\documentclass{article}
\usepackage{spconf,amsmath,graphicx}
\usepackage{longtable}
\usepackage{enumitem}
\usepackage{hyperref}
\usepackage{graphicx}
\usepackage{booktabs}
\usepackage{makecell}
\usepackage{multirow}
\usepackage{array}
\usepackage{tabularx}
\setlist{nosep, leftmargin=14pt}
\usepackage{float}
\usepackage{mwe} 

\usepackage{amsmath} 

\title{XLSTM-HVED: Cross-Modal Brain Tumor Segmentation and MRI Reconstruction Method Using Vision XLSTM and Heteromodal Variational Encoder-Decoder}
\name{
\parbox{\linewidth}{\centering
Shenghao Zhu$^1$, Yifei Chen$^1$, Shuo Jiang$^1$, Weihong Chen$^1$, Chang Liu$^1$,\\ \textit{Yuanhan Wang}$^1$, \textit{Xu Chen}$^1$, \textit{Yifan Ke}$^1$,  \textit{Feiwei Qin}$^{1,\star}$, \textit{Changmiao Wang}$^2$, \textit{Zhu Zhu}$^{3,4,\star}$}
}
\address{$^1$Hangzhou Dianzi University, Hangzhou, China \\
$^2$Shenzhen Research Institute of Big Data, Shenzhen, China \\
$^3$Children's Hospital, Zhejiang University School of Medicine, Hangzhou, China \\
$^4$Sino-Finland Joint AI Laboratory for Child Health of Zhejiang Province, Hangzhou, China \\
\tt$\!\!\!\star$ Corresponding Author: zhuzhu$\_$cs@zju.edu.cn, qinfeiwei@hdu.edu.cn\\}

\begin{document}
%
\maketitle
\begin{abstract}
Neurogliomas are among the most aggressive forms of cancer, presenting considerable challenges in both treatment and monitoring due to their unpredictable biological behavior. Magnetic resonance imaging (MRI) is currently the preferred method for diagnosing and monitoring gliomas. However, the lack of specific imaging techniques often compromises the accuracy of tumor segmentation during the imaging process. To address this issue, we introduce the XLSTM-HVED model. This model integrates a hetero-modal encoder-decoder framework with the Vision XLSTM module to reconstruct missing MRI modalities. By deeply fusing spatial and temporal features, it enhances tumor segmentation performance. The key innovation of our approach is the Self-Attention Variational Encoder (SAVE) module, which improves the integration of modal features. Additionally, it optimizes the interaction of features between segmentation and reconstruction tasks through the Squeeze-Fusion-Excitation Cross Awareness (SFECA) module. Our experiments using the BraTS 2024 dataset demonstrate that our model significantly outperforms existing advanced methods in handling cases where modalities are missing. Our source code is available at \href{https://github.com/Quanato607/XLSTM-HVED}{\text{https://github.com/Quanato607/XLSTM-HVED}}.
\end{abstract}

\vspace{-1.0em}
\section{Introduction}
\label{sec:intro}

Neurogliomas rank among the deadliest cancers and are the most common malignant primary brain tumors in adults, accounting for approximately 25\% of all primary brain tumors and 80\% of malignant primary brain and central nervous system tumors. Within this category, diffuse gliomas are the most frequently occurring malignant subtype. However, their diverse biological behavior and unpredictable response to treatment present substantial challenges for treatment \cite{sharma2019missing}.

Magnetic resonance imaging (MRI) is the preferred method for diagnosing diffuse gliomas, offering crucial insights into tumor size, location, and morphology over time \cite{ge2024tc}. Various MRI modalities, such as T1, T1-Gd, T2, and FLAIR, provide complementary information about brain tumors. For instance, T1 and T2 modalities help identify vasogenic edema in subacute strokes; T1-Gd, enhanced with contrast, reveals details about vasculature and the blood-brain barrier; and FLAIR provides general information about stroke lesions. Consequently, using multimodal MRI for segmentation reduces uncertainty and enhances performance compared to relying on single modalities. However, in practice, missing modalities are common. This issue arises due to patients' limited tolerance for scans, time constraints, or image corruption. As a result, accurate MRI segmentation in the presence of missing modalities has garnered significant research interest in recent years. For instance, Doren et al. \cite{dorent2019hetero} developed a Heteromodal Variational Encoder-Decoder (HVED) for tumor segmentation and modal complementation under conditions of missing modalities. Similarly, Chen et al. \cite{chen2019robust} proposed a multimodal brain segmentation method based on a feature decoupling and gated fusion framework to enhance robustness when modalities are absent. Furthermore, Liu et al. \cite{liu2023m3ae} introduced a multimodal mask autoencoder (M3AE) for brain tumor segmentation in scenarios where multiple modalities are missing.


\begin{figure*}[htb]
\begin{minipage}[b]{1.0\linewidth}
  \centering
  \centerline{\includegraphics[width=0.90\textwidth]{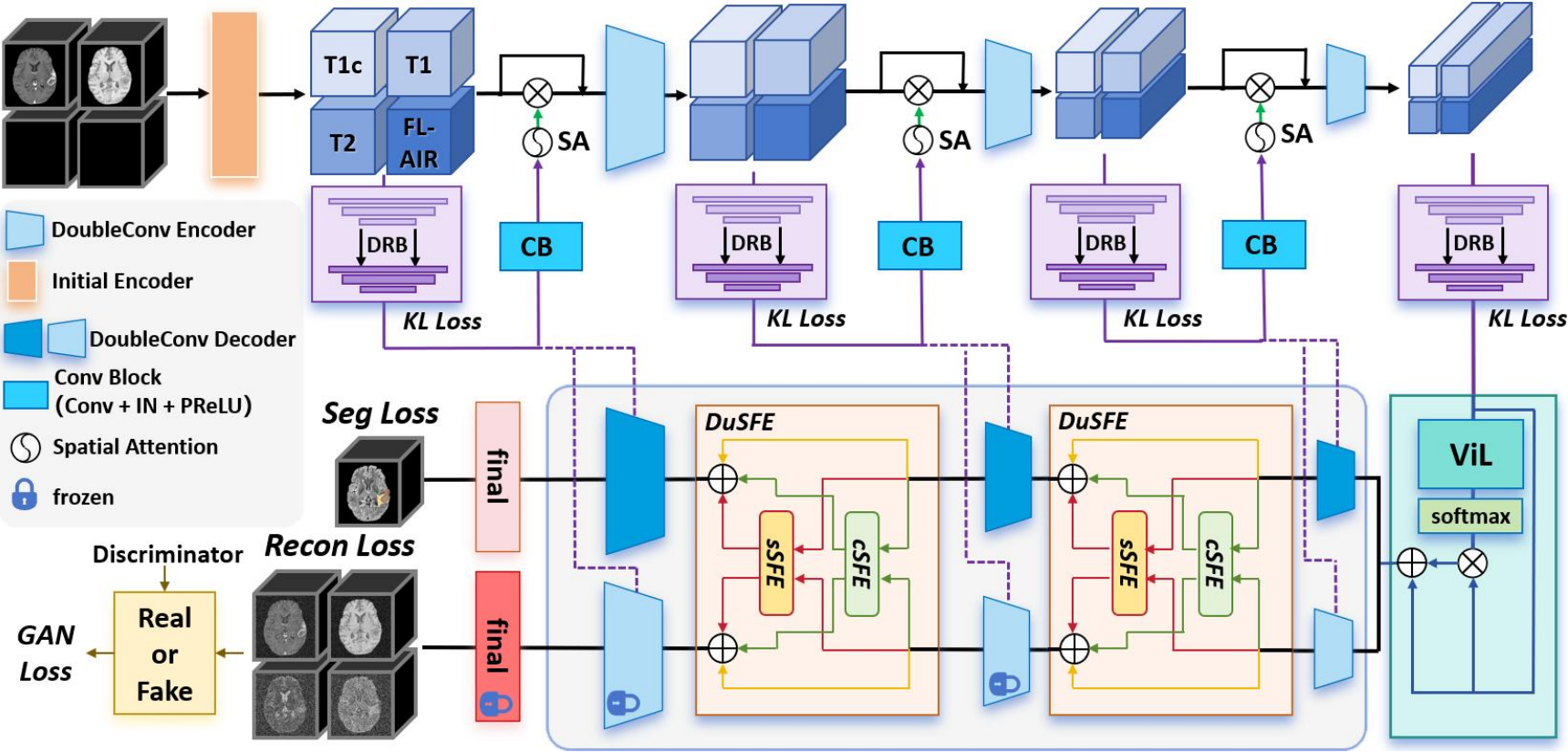}}
\end{minipage}
\vspace{-10pt}
\caption{The overall framework of the proposed XLSTM-HVED model. The model comprises the Self-Attention Variational Encoder, Vision XLSTM Attention Module, and Squeeze-Fusion-Excitation Cross Awareness Module.}
\label{fig:1}
\vspace{-10pt}  
\end{figure*}

\vspace{-1.0em}
\section{Methods}
\label{sec:methods}
\vspace{-1.0em}
\subsection{Model Overview}
\label{ssec:overview}

As illustrated in Figure \ref{fig:1}, we introduce the XLSTM-HVED model, designed to improve the interaction of features between different MRI modalities. This model employs the Heteromodal Variational Encoder-Decoder (HVED) to capitalize on the capabilities of XLSTM. It merges channel and spatial information through the SFECA module, effectively utilizing complementary information from various modalities. Consequently, the model can reconstruct existing MRI data, supplement missing data, and accurately segment tumor lesions, achieving synchronization across multiple tasks seamlessly.

\vspace{-1.0em}
\subsection{Self-Attention Variational Encoder}
\label{sssec:SAVE}

As shown in Figure \ref{fig:1}, our proposed Self-Attention Variational Encoder (SAVE) encodes multimodal inputs using the Multimodal Variational Autoencoder (MVAE) \cite{wu2018multimodal}. The encoded features are processed through a convolutional layer, followed by spatial attention, to enhance focus on relevant multimodal features. These encoded features are simultaneously linked to the corresponding decoder. Developed under the framework of conditionally independent modalities \(X\, =\, x_1, \ldots, x_n\) given a shared latent variable \(z\), MVAE extends the Variational Autoencoder (VAE) \cite{kingma2013auto} to manage missing data by incorporating multimodal inputs.

The module fuses the means \(\mu\) and covariances \(\Sigma\) of each modality using a Product of Gaussians (PoG) \cite{cao2014generalized} to form a latent variable \(z\), effectively excluding any missing modalities. The sampled \(z\) is then decoded into image space via the reparameterization trick, with feature representation further refined by the Dimension Reduction Block (DRB) \cite{jeong2022region}. In the final downsampling layer, multimodal features are merged with unimodal latent features using MVAE and DRB, and are subsequently fed into the Vision-LSTM Attention (ViLA) module. This integration ensures that encoded multimodal features are effectively combined with latent variable features derived from unimodal encoding.

\begin{figure*}[htb]
\begin{minipage}[b]{1.0\linewidth}
  \centering
  \centerline{\includegraphics[width=0.90\textwidth]{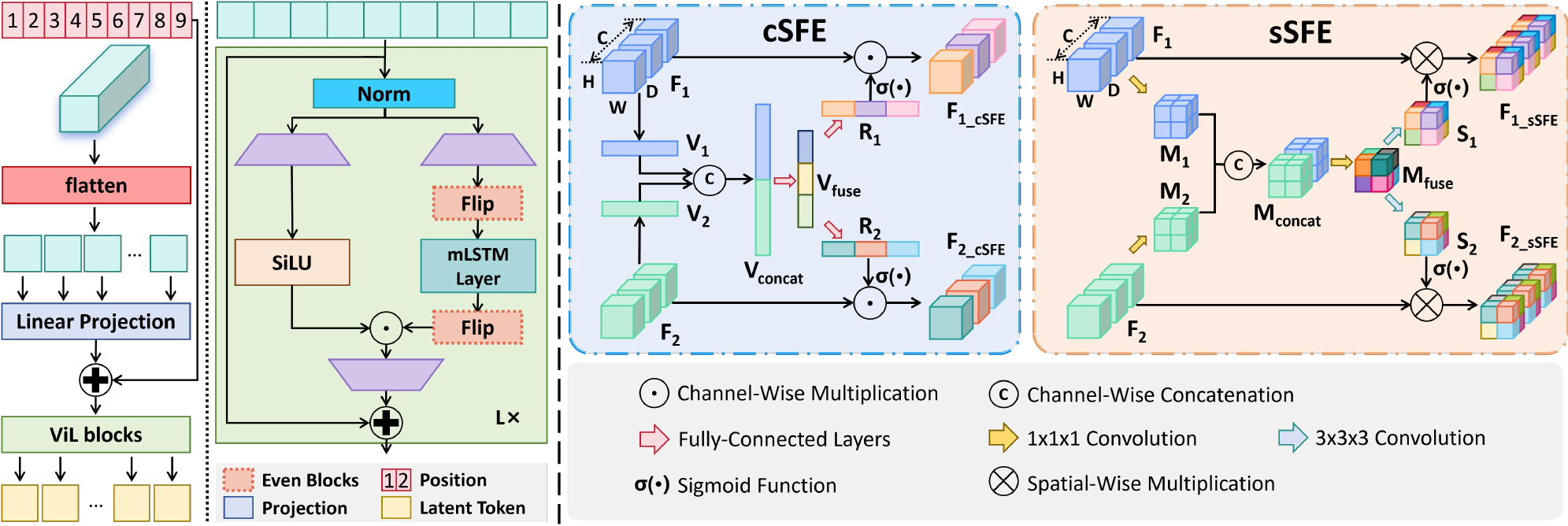}}
\end{minipage}
\vspace{-10pt}
\caption{ \textbf{(left)} Detailed structure of the Vision-LSTM module. \textbf{(right)} Detailed structure of the cSFE and sSFE modules.}
\label{fig:2}
\vspace{-10pt}  
\end{figure*}

\vspace{-1.0em}
\subsection{Vision XLSTM Attention Module}
\label{sssec:xlstm}
As shown in Figure \ref{fig:1}, our proposed ViLA module processes features through the Vision-LSTM (ViL) module \cite{alkin2024vision} and applies attention to the original input using a softmax operation. The features, shaped as (B, C, H, W, D), are obtained from the downsampling in the SAVE module and are reshaped to (B, H*W*D, C) before being fed into the ViL module. The detailed structure of this module is depicted in Figure \ref{fig:2} (left). Within the ViL module, the input feature map, originally of shape (B, H*W*D, C), is split into non-overlapping tokens of size H*W*D using a shared linear projection. This division facilitates detailed processing of each segment. Learnable positional embeddings are then added to each token to enhance the model’s spatial awareness. Central to the ViL module are the mLSTM blocks, which integrate matrix memory and apply covariance update rules to improve feature representation and processing efficiency.

\vspace{-1.0em}
\subsection{Squeeze-Fusion-Excitation Cross Awareness Module}
\label{sssec:dual-branch}

The features generated by the ViLA module are directed into both the segmentation and reconstruction decoders. Concurrently, the downsampled features from the SAVE module are integrated into the decoding process. During upsampling, the 8- and 16-dimensional features from the decoders interact through the DuSFE module \cite{chen2023dusfe}, effectively merging their outputs for final feature extraction tailored to different tasks.

In the pre-training phase, only the reconstruction decoder is actively trained, while layers 2, 3, and the final module are kept static. The DuSFE module plays a crucial role in combining segmentation and reconstruction features. It employs channel-squeeze-fusion-excitation (cSFE) and spatial-squeeze-fusion-excitation (sSFE) techniques for spatial recalibration, as illustrated in Figure \ref{fig:2} (right). Within cSFE, features \(F_1\) and \(F_2\) are compressed into vectors \(V_1\) and \(V_2\) using global average pooling. These vectors are then concatenated and processed through fully connected layers to produce a fusion vector \(V_{fuse}\). This fusion vector is utilized to derive \(R_1\) and \(R_2\), which, through channel multiplication, generate \(F_{1\_cSFE}\) and \(F_{2\_cSFE}\). For spatial recalibration, sSFE begins by processing \(F_1\) and \(F_2\) with a convolutional layer. The resultant features are concatenated and further processed by an additional convolutional layer to form \(M_{fuse}\). This fused feature map is then introduced into two separate convolutional pathways, where spatial multiplication is applied to obtain the refined outputs \(F_{1\_sSFE}\) and \(F_{2\_sSFE}\).


\vspace{-1.0em}
\section{EXPERIMENT RESULT}
\vspace{-1.0em}
\subsection{Dataset and Implementation Details}
\label{sec:data}

Our study utilizes the multimodal Brain Tumor Segmentation Challenge (BraTS) 2024 dataset \cite{de20242024}. This dataset includes MRI scans from T1, T1ce, T2, and FLAIR modalities of approximately 4,500 patients diagnosed with various gliomas. The images have been resampled to a uniform one mm³ resolution, aligned to anatomical templates, and pre-processed to remove cranial structures. Expert radiologists manually labeled the images, focusing segmentation on three key regions.

For our experiments, we divided the dataset randomly into training and testing sets, comprising 80\% and 20\% of the total data, respectively. We conducted our experiments using the PyTorch framework, leveraging an NVIDIA Tesla V100 for computational support. During training, we utilized a batch size of 2, set a learning rate of 0.0001, and carried out the training over a period of 72 hours.

\begin{table*}[htbp]
\caption{Comparison of HD95 and Dice score for various state-of-the-art models utilizing the BraTS 2024 dataset.}
\centering
\setlength{\tabcolsep}{0.2mm}{
\begin{tabular}{c|c|c|c|cccc|cccc|cccc}
    \hline
    \multicolumn{4}{c|}{\textbf{\makecell[c]{Avaliable\\Modalities}}} & \multicolumn{4}{c|}{\textbf{Whole tumor}} & \multicolumn{4}{c|}{\textbf{Tumor core}} & \multicolumn{4}{c}{\textbf{Enhancing tumor}} \\
    \hline
    \multicolumn{16}{c}{\textbf{Score of HD95 (mm) $\downarrow$}} \\
    \hline
    FL & T1 & T1c & T2 & \makecell[c]{RA-\\HVED} & \makecell[c]{RMBTS} & \makecell[c]{mmformer} & Ours & \makecell[c]{RA-\\HVED} & \makecell[c]{RMBTS} & \makecell[c]{mmformer}  & Ours & \makecell[c]{RA-\\HVED} & \makecell[c]{RMBTS} & \makecell[c]{mmformer} & Ours\\
    \hline
    $\circ$ & $\circ$ & $\circ$ & $\bullet$ & 22.12& 39.05
& 19.45& \textbf{17.07} & 25.32& 24.83& 27.70& \textbf{16.06} & \textbf{12.85}& 23.79
& 26.36& 18.57\\ 
    \hline
    $\circ$ & $\circ$ & $\bullet$ & $\circ$ & 40.16& 63.59
& 51.96& \textbf{27.33} & 30.41& 23.11& 62.12& \textbf{21.27} & 24.96& \textbf{21.90}& 59.83& 23.68\\
    \hline
    $\circ$ & $\bullet$ & $\circ$ & $\circ$ & 57.69& 57.68
& 40.66& \textbf{37.39} & 57.06& 47.09
& 39.07& \textbf{33.81} & 46.98& 44.79& 37.55& \textbf{24.67}\\
    \hline
    $\bullet$ & $\circ$ & $\circ$ & $\circ$ & 23.75& 59.42
& \textbf{18.21}& 21.36 & \textbf{22.50}& 24.09& 24.29& 22.57 & 15.15& 23.73
& 23.22& \textbf{12.65}\\ 
    \hline
    $\circ$ & $\circ$ & $\bullet$ & $\bullet$ & 19.80& 36.05
& 18.80& \textbf{12.27}& 15.77& 19.76
& 25.63& \textbf{11.95} & 14.89& 19.16
& 24.02& \textbf{10.81}\\
    \hline
    $\circ$ & $\bullet$ & $\bullet$ & $\circ$ & 34.78& 50.11
& 34.53& \textbf{23.36} & 26.77& 25.76
& 38.72& \textbf{23.39} & 23.69& 24.18
& 36.72& \textbf{13.98}\\
    \hline
    $\bullet$ & $\bullet$ & $\circ$ & $\circ$ & 20.91& 41.70& \textbf{13.94}& 23.38 & 20.93& 23.73& \textbf{19.67} &20.18 & \textbf{13.24}& 22.40& 18.62& 16.07\\ 
    \hline
    $\circ$ & $\bullet$ & $\circ$ & $\bullet$ & 17.35& 33.12
& 16.87& \textbf{14.45} & 23.11& 21.89& 24.13& \textbf{13.27} & \textbf{10.94}& 21.90& 22.17& 15.04\\
    \hline
    $\bullet$ & $\circ$ & $\circ$ & $\bullet$ & 16.94& 37.44
& 13.13& \textbf{12.61} & 19.66& 19.11
& 19.28& \textbf{12.11} & \textbf{10.83}& 19.54
& 18.43& 17.17\\
    \hline
    $\bullet$ & $\circ$ & $\bullet$ & $\circ$ & 21.23& 47.78
& \textbf{15.51}& 17.66 & \textbf{15.88}& 18.51
& 20.48& \textbf{17.32} & 14.01& 17.24
& 18.34& \textbf{10.91}\\ 
    \hline
    $\bullet$ & $\bullet$ & $\bullet$ & $\circ$ & 20.52 & 34.77
& \textbf{13.35}& 16.88 & \textbf{14.36} & 16.30& 17.30& 16.47 & 14.23 & 15.08
& 16.43& \textbf{10.05}\\ 
    \hline
    $\bullet$ & $\bullet$ & $\circ$ & $\bullet$ & 14.98 & 33.23
& 12.94& \textbf{11.96} & 21.59 & 20.04
& 18.71& \textbf{11.30} & 11.04 & 19.46
& 17.66& \textbf{9.87}\\
    \hline
    $\bullet$ & $\circ$ & $\bullet$ & $\bullet$ & 16.31 & 35.32
& 12.22&  \textbf{10.92} & 13.34 & 15.55
& 15.37&\textbf{10.47} & 12.83 & 15.24
& 14.50& \textbf{9.76}\\ 
    \hline
    $\circ$ & $\bullet$ & $\bullet$ & $\bullet$ & 18.58& 34.11
& 16.81& \textbf{12.57} & 16.16& 14
.00&22.08& \textbf{11.99} & 15.40& 13.54
& 20.36& \textbf{9.70}\\
    \hline
    $\bullet$ & $\bullet$ & $\bullet$ & $\bullet$ & 15.93 & 34.02
& 11.81& \textbf{11.73} & 12.49 & 13.74
& 14.73& \textbf{11.30} & 12.18 & 13.33
& 13.95& \textbf{8.74}\\
    \hline
    \multicolumn{4}{c|}{\textbf{Average}} & 24.07 & 42.49& 20.68& \textbf{17.86} & 22.36 & 21.83& 25.95& \textbf{16.90} & 16.88 & 20.95&24.54& \textbf{14.11}\\ 
    \hline
    \multicolumn{16}{c}{\textbf{Score of Dice (\%) $\uparrow$}} \\
    \hline
        FL & T1 & T1c & T2 & \makecell[c]{RA-\\HVED} & \makecell[c]{RMBTS} & \makecell[c]{mmformer}  & Ours & \makecell[c]{RA-\\HVED} & \makecell[c]{RMBTS} & \makecell[c]{mmformer}  & Ours & \makecell[c]{RA-\\HVED} & \makecell[c]{RMBTS} & \makecell[c]{mmformer}  & Ours\\
    \hline
    $\circ$ & $\circ$ & $\circ$ & $\bullet$ & 75.44& 70.06
& 72.60 & \textbf{77.23} & 26.47& 10.92
& \textbf{47.24}& 36.60 & 35.79& 7.94
& 44.90 & \textbf{46.60}\\ 
    \hline
    $\circ$ & $\circ$ & $\bullet$ & $\circ$ & 51.34& 51.22
& 55.45&  \textbf{60.25} & 54.23 & 36.52
& 52.29& \textbf{66.28} & 37.80& 37.83
& 50.48& \textbf{50.55}\\
    \hline
    $\circ$ & $\bullet$ & $\circ$ & $\circ$ & 9.48& 51.75
& \textbf{61.32}& 31.42 & 9.42& 12.57
& \textbf{44.39}& 23.23 & 9.24& 10.01
& \textbf{42.27}& 36.84\\
    \hline
    $\bullet$ & $\circ$ & $\circ$ & $\circ$ & 71.42& 65.05
& 72.68 & \textbf{73.88} & 41.09& 11.20& 33.10& \textbf{47.53} & 39.81& 8.16
& 31.36& \textbf{47.28}\\ 
    \hline
    $\circ$ & $\circ$ & $\bullet$ & $\bullet$ & 77.50& 75.25
& 74.28& \textbf{82.35} & 61.30& 40.43
& 62.60& \textbf{72.43} & 42.30& 41.89
& \textbf{61.33}& 55.28\\
    \hline
    $\circ$ & $\bullet$ & $\bullet$ & $\circ$ & 53.43& 60.57
& \textbf{65.36}& 60.29 & 54.80& 37.60& 60.58& \textbf{66.57} & 36.59& 40.10& \textbf{58.96}& 52.11\\
    \hline
    $\bullet$ & $\bullet$ & $\circ$ & $\circ$ & 72.94& 76.40& \textbf{79.20} & 75.82 & 41.89& 16.79
& \textbf{49.58}& 47.74 & 42.55& 13.10& 45.28& \textbf{48.63}\\ 
    \hline
    $\circ$ & $\bullet$ & $\circ$ & $\bullet$ & 76.06& 74.95
& 75.12 & \textbf{78.16} & 29.19& 15.23
& \textbf{51.11}& 37.07 & 43.82& 11.80&49.36& \textbf{50.16}\\
    \hline
    $\bullet$ & $\circ$ & $\circ$ & $\bullet$ & 80.09& 77.31
& 79.64& \textbf{83.19} & 40.48& 14.48
& 49.62& \textbf{50.43} & 44.42& 10.80& 46.60& \textbf{53.35}\\
    \hline
    $\bullet$ & $\circ$ & $\bullet$ & $\circ$ & 72.85& 76.03
& \textbf{78.27}& 76.94 & 61.93& 38.86
& 60.55& \textbf{72.78} & 44.11& 40.64
& \textbf{59.33}& 55.87\\ 
    \hline
    $\bullet$ & $\bullet$ & $\bullet$ & $\circ$ & 72.91 & 79.33
& 79.97 & \textbf{80.91} & 62.48 & 40.13
& 64.32& \textbf{72.95} & 43.92 & 43.49
& \textbf{62.96}& 58.94\\ 
    \hline
    $\bullet$ & $\bullet$ & $\circ$ & $\bullet$ & 80.63 & 79.71
& 80.71 & \textbf{83.65} & 43.24 & 17.40& \textbf{52.64}& 51.99 & 48.40 & 13.99
& 49.59& \textbf{56.22}\\
    \hline
    $\bullet$ & $\circ$ & $\bullet$ & $\bullet$ & 80.40 & 80.29& 81.01 & \textbf{84.80} & 64.00 & 40.42
& 65.52&\textbf{76.34} & 46.87 & 42.34
& \textbf{63.60}& 59.11\\ 
    \hline
    $\circ$ & $\bullet$ & $\bullet$ & $\bullet$ & 77.73& 76.12
& 75.57 &\textbf{82.22} & 61.85& 40.87
& 65.27& \textbf{72.96} & 40.67& 44.14
& 64.16& \textbf{64.22}\\
    \hline
    $\bullet$ & $\bullet$ & $\bullet$ & $\bullet$ & 80.09 & 80.92
& 81.27  & \textbf{86.77} & 64.97 & 40.61
& 67.01& \textbf{77.91} & 45.88 & 55.22
& 65.73& \textbf{65.90}\\
    \hline
    \multicolumn{4}{c|}{\textbf{Average}} & 68.82 & 71.66& 74.16& \textbf{74.53} & 47.82 & 27.60& 55.05& \textbf{58.19} & 39.77 & 27.36&52.99& \textbf{53.83}\\ 
    \hline
    
\end{tabular}}
\label{table2}
\end{table*}

\vspace{-1.0em}
\subsection{Comparative Experiment}
For evaluation, we utilized the RA-HVED \cite{jeong2022region} and RMBTS \cite{chen2019robust} models as baselines. All the baselines and our model were trained and tested using the same backbone network to ensure consistency in the evaluation phase. As detailed in Table \ref{table2}, our model's performance is compared to these baselines using Dice scores and HD95 scores. We selected U-HVED \cite{dorent2019hetero} as the backbone network. Our model demonstrates strong performance across all metrics and is particularly effective in scenarios involving multimodal fusion. Furthermore, we compared the XLSTM-HVED model to several baseline models and the state-of-the-art. Table \ref{table2} highlights that the XLSTM-HVED model achieves competitive results in the HD95 metric compared to other advanced models like mmformer \cite{zhang2022mmformer}, particularly in multimodal fusion contexts. However, it is noted that the XLSTM-HVED model has lower Dice metrics for Enhancing tumor segmentation.

\vspace{-1.0em}
\subsection{Ablation Study}
We investigated the impact of the SAVE, ViLA, and SFECA modules on model performance. The findings are summarized in Table \ref{table4}. Removing the ViLA module led to a noticeable decline in segmentation performance for the T1 modality, with both HD95 and Dice metrics worsening. Similarly, excluding the SAVE module reduced the overall performance of multimodal fusion, particularly affecting the segmentation of the whole tumor. Omitting the SFECA module resulted in significant degradation in segmentation performance for the T1c and T1 modalities, especially in the HD95 metrics. These results underscore the importance of these three modules.


\begin{table*}[htbp]
\caption{Comparison of average HD95 and Dice score for ablation experiments utilizing the BraTS 2024 dataset.}
\centering
\setlength{\tabcolsep}{5.5mm}{
\begin{tabular}{c|c|c|c|c|c|c}
    \hline
    \multirow{2}*{\centering Method} & \multicolumn{3}{c|}{\textbf{Average Dice Score (\%) $\uparrow$}} & \multicolumn{3}{c}{\textbf{Average HD95 Score (mm) $\downarrow$}} \\
    \cline{2-7}
    ~ & \centering WT & \centering TC & \centering ET & \centering WT & \centering TC & ET \\
    \hline
    w/o SAVE & 66.25 & 51.66 & 45.05 & 28.74 & 28.18 & 30.47 \\
    \hline
    w/o ViLA Module & 65.05 & 52.34 & 44.97 & 22.69 & 22.36 & 16.41 \\
    \hline
    w/o SFECA Module & 68.11 & 54.34 & 39.97 & 46.39 & 41.17 & 33.01 \\
    \hline
    XLSTM-HVED (Ours) & \textbf{74.53} & \textbf{58.19} & \textbf{53.83} & \textbf{17.86} & \textbf{16.90} & \textbf{14.11} \\
    \hline
\end{tabular}}
\label{table4}
\end{table*}

\vspace{-1.0em}
\section{Conclusion}
\label{sec:page}

In this paper, we introduce the XLSTM-HVED model to address brain tumor segmentation challenges, particularly in scenarios where some MRI modalities are unavailable. Our model enhances segmentation accuracy and MRI data reconstruction quality by integrating cross-modal encoding, multi-task learning, and attention mechanisms. The SAVE module facilitates effective multimodal information fusion, while the SFECA module optimizes the coordination between segmentation and reconstruction tasks. Our extensive experiments on the BraTS 2024 dataset reveal that the model is highly robust across various tasks, even when dealing with the loss of certain modalities. Additionally, ablation studies confirm that removing any module significantly diminishes the model's overall performance, highlighting the critical role each component plays in ensuring the model's effectiveness.


\clearpage

\section{Compliance with Ethical Standards}
This research study was conducted retrospectively using human subject data made available in open access by BraTS 2024 dataset \cite{de20242024}. Ethical approval was not required as confirmed by the license attached with the open access data.

\section{Acknowledgment}
This work was supported by Key R\&D Program of Zhejiang (2023C03101), Medical Health Science and Technology Project of Zhejiang Provincial Health Commission (2023KY832), the Open Project Program of the State Key Laboratory of CAD\&CG, Zhejiang University (No. A2304), Guangdong Basic and Applied Basic Research Foundation (No. 2022A1515110570), Guangxi Key R\&D Project (No. AB24010167), Innovation Teams of Youth Innovation in Science, Technology of High Education Institutions of Shandong Province (No.2021KJ088) and Shenzhen Science and Technology Program (No. KCXFZ20201221173008022), Shenzhen Stability Science Program 2022 (2023SC0073).


\end{document}